\documentclass[aps,prb, superscriptaddress, noeprint, twocolumn]{revtex4-2}  %
\usepackage{bbold}
\usepackage{comment}
\usepackage[colorlinks, citecolor=red]{hyperref}
\usepackage{hyperref}
\usepackage[utf8]{inputenc}
\usepackage[T1]{fontenc}    %
\usepackage[english]{babel}
\usepackage[pdftex]{graphicx}
\usepackage{amsmath,amssymb,amsthm,bbm, mathtools} %
\usepackage{mathrsfs}
\usepackage{xcolor}
\usepackage[normalem]{ulem}
\usepackage{placeins}

\DeclareMathSymbol{\shortminus}{\mathbin}{AMSa}{"39}
\renewcommand{\vec}{\boldsymbol}

\begin{document}

\title{Spin-qubit readout analysis based on a hidden Markov model}

\author{Maria Spethmann}
\email{maria.spethmann@unibas.ch}
\affiliation{Department of Physics, University of Basel, 4056 Basel, Switzerland}
\affiliation{RIKEN, Center for Emergent Matter Science (CEMS), Wako-shi, Saitama 351-0198, Japan}
\author{Peter Stano}
\email{peter.stano@riken.jp}
\affiliation{RIKEN, Center for Emergent Matter Science (CEMS), Wako-shi, Saitama 351-0198, Japan}
\author{Daniel Loss}
\affiliation{Department of Physics, University of Basel, 4056 Basel, Switzerland}
\affiliation{RIKEN, Center for Emergent Matter Science (CEMS), Wako-shi, Saitama 351-0198, Japan}

\begin{abstract}
Across most qubit platforms, the readout fidelities do not keep up with the gate fidelities, and new ways to increase the readout fidelities are searched for.
For semiconductor spin qubits, a typical qubit-readout signal consists of a finite stretch of a digitized charge-sensor output. Such a \textit{signal trace} is usually analyzed by compressing it into a single value, either maximum or sum. 
The binary measurement result follows by comparing the single value to a decision threshold fixed in advance. This \textit{threshold method}, while simple and fast, omits information that could potentially improve the readout fidelity. Here, we analyze what can be achieved by more sophisticated signal-trace processing using the hidden Markov model (HMM). The HMM is a natural choice, being the optimal statistical processing if the noise is white. It also has a computationally efficient implementation, known as the forward-backward algorithm, making HMM processing practical.
However, unlike in many computer-simulation studies, in real experiments the noise is correlated. How this change affects the HMM implementation and reliability is our subject. We find that the HMM using white noise as the system statistical model is surprisingly sensitive to correlations; it only tolerates very small correlation times. We suggest alleviating this deficiency by a signal prefiltering. The correlations have a similar strongly negative impact on the HMM model calibration (the Baum-Welch algorithm). Besides studying the effects of noise correlations, as a specific application of the HMM we calculate the readout fidelity at elevated temperatures, relevant to recent experimental pursuits of hot spin qubits.

\end{abstract}

\maketitle

\section{Introduction}

Spin qubits in semiconductor quantum dots are promising candidates to build a digital quantum computer, standing out with long coherence times, full electrical control, and the possibility to scale up using the well-developed fabrication techniques of the semiconductor industry \cite{Loss1998, Hanson2007, Burkard2023}. Recent advances include  arrays of electron and hole spin qubits in silicon and germanium with high-fidelity single- and two-qubit gates \cite{Bulaev2005, Philips2022, John2025, Mills2022b, Hendrickx2021, Liles2024, George2024,  Stano2022}. 
The qubit-readout speed and fidelity are critical for quantum error correction \cite{Hetenyi2024}. However, simultaneously meeting the demands on qubit coherence times, qubit number, connectivity,  gate fidelity, and readout fidelity is a complex challenge with unavoidable trade-offs. Once the experimental setting is fixed, it is crucial that the available measurement data is analyzed optimally, extracting as much information as possible \cite{Gambetta2007}.

The two common ways to read out a spin qubit are Pauli spin blockade (PSB) readout \cite{Ono2002, Oakes2023, HarveyCollard2018, Seedhouse2021} and Elzerman readout \cite{Elzerman2004, Oakes2023,Keith2019, Mills2022}. For both readouts, the qubit outcome is often inferred by first compressing the charge-sensor signal trace into a single value, the integrated signal in the case of PSB readout \cite{Takeda2024} and the peak signal in the case of Elzerman readout \cite{Keith2019, Mills2022, Gorman2017}. This value is then compared to a threshold value. Inferring the qubit state from this threshold method is convenient since it is fast and simple. However, it raises the question of how much information has been lost by compressing a signal trace into a single value and how that impacts the readout fidelity \cite{Gambetta2007,DAnjou2014}.

The readout fidelity is ultimately limited by sensor noise and the stochastic nature of transitions among qubit states \cite{HarveyCollard2018, Takeda2024,Keith2019,DAnjou2014}. In this article, we investigate if more sophisticated processing of the readout signal improves the readout fidelity. We describe the qubit measurement by a hidden Markov model (HMM). Our basic motivation is that it delivers the exact conditional distribution (probability of the qubit state, given the observed signal trace) for systems with white noise (that is, noise with negligibly short autocorrelation time) without restricting qubit state transitions during the measurement in any way.

Besides being the exact and optimal solution in the well-motivated limit of white noise, HMM also has further appealing features. First, the exact conditional probabilities can be evaluated by the forward-backward algorithm, which scales linearly with the signal length \cite{Rabiner1989}. Thus, HMM has the same computational complexity as the threshold method. Second, the forward-backward algorithm allows for recursive updates of the conditional probabilities upon the arrival of a new data point \cite{Xue2020}. This feature makes the HMM suitable for an online processing that allows the dynamic adjustment of the measurement time to reach a desired fidelity, and thus minimizes the average readout time for a given readout fidelity \cite{DAnjou2016}.  Third, the model parameters can be learned automatically using the so-called Baum-Welch algorithm \cite{House2009, House2013}, without any separate model calibrations. The algorithm only requires training measurement data, produced by repeating the exact same measurement protocol (say, on random qubit input) as it is used in actual quantum-circuit implementations. Finally, as the forward-backward algorithm provides the qubit state probabilities rather than hard readout decisions, the HMM output is suitable for soft decoding \cite{DAnjou2014b, Xue2020, DAnjou2021}, which means boosting fidelity by combining several readouts \cite{Guizzo2004}. 
 
However, the HMM benefits just discussed have the following context. The idealized setup under which the HMM is the exact solution assumes that the noise is white and that the statistical model of the actual system is exactly known and exactly implemented within the forward-backward algorithm. This is never so, and the difference between the true and adopted model can destroy the gains from the more elaborate data-processing methods. We focus on correlated noise, prevalent in solid-state implementations. We examine how much correlation the HMM can tolerate before it breaks down, both in terms of parameter estimation (Baum-Welch algorithm) and readout fidelity (forward-backward algorithm). We find that both algorithms are instable with respect to noise correlations. For the specific model of a Gaussian correlation function that we choose, correlation times as short as a few steps are already fatal for the HMM method, for both its calibration and the application, which reaches inferior fidelity. For this specific noise correlation, we can reestablish the HMM algorithm's superiority by a signal prefiltering that reduces the correlations.

We also analyze the Elzerman readout, a more complex readout scheme that involves three states instead of two. We again first study how the HMM fidelity is affected by the parameter calibration, after which we examine readout at finite temperature. Here, the combined effects of thermal excitations and noise make it difficult to calculate the optimal fidelity using alternative methods. The HMM enables us to map out how the readout fidelity transitions from a low-temperature to a high-temperature regime.
 
The idea that an improved analysis method of the measurement data could raise the readout fidelity is not new. Gambetta \textit{et al.} (2007) \cite{Gambetta2007} compared different filters to analyze the measurement records of a decaying qubit and they found an optimal nonlinear filter that provides the best possible readout fidelity. D’Anjou and Coish (2014) \cite{DAnjou2014} differentiated between a deterministic turn-on time (suited to PSB readout) and a stochastic turn-on time (suited to Elzerman readout), found an optimal analysis for both, and compared them with different threshold methods. We will review and confirm some of their results using the HMM as an introduction to the topic and build upon the existing literature.
House \textit{et al.} (2013) \cite{House2013} did not calculate readout fidelities but use an HMM to estimate the spin-dependent dot-lead tunnel rates at elevated temperatures, including confidence intervals.
Harvey-Collard \textit{et al.} (2018) \cite{HarveyCollard2018} successfully applied a maximum likelihood analysis to PSB readout data that captured time correlations exactly but neglected qubit transitions. Further alternatives to postprocess qubit measurement data  are  the change point detection theory \cite{Mizokuchi2020} or the wavelet edge detection \cite{Prance2015}. They aimed at detection of large changes in a signal which thus could also be robust to long-time correlations. Nevertheless, we choose to benchmark the HMM to the threshold method as the current standard.

The article is structured as follows. In Sec. \ref{sec:white_noise}, we introduce the HMM and explain how we use the forward-backward algorithm to assign the most probable qubit state. We apply that algorithm to simulated signal traces with white noise and, in line with previous literature, demonstrate that the threshold method is suboptimal. We choose the PSB readout here as a scheme that is simpler and more easily transferable to other qubit platforms. In Sec. \ref{sec:correlated_noise}, we change the noise from white to correlated and test how much correlation the HMM method can tolerate before it becomes worse than the threshold method.  
We move to the Elzerman readout in Sec. \ref{sec:Elzerman_fidelity}. Here we calculate readout fidelities with HMM parameters that were calibrated with the Baum-Welch algorithm and analyze how temperature affects the readout fidelities. We conclude in Sec. \ref{sec:conclusion}. 
In the Appendixes, we shortly describe a scenario in which the threshold method is optimal, we review the Baum-Welch algorithm to estimate HMM parameters, review the generation of correlated noise signals from a given spectral density, and examine how the Baum-Welch parameter estimation is impacted by correlated noise.

\section{White noise}
\label{sec:white_noise}

In this section, we compare the threshold and HMM methods on computer-simulated readout data with white noise. 
We pick PSB readout as the simplest case: There are two qubit states to tell apart, one of which can decay to the other during the measurement. This scenario applies to several other qubit architectures.
The purpose of this section is threefold: (1) We introduce the HMM as a qubit-readout model and the forward-backward algorithm as the evaluation of the qubit-state probabilities conditioned on the observed signal. (2) In line with previous literature \cite{Gambetta2007, DAnjou2014}, we demonstrate that the threshold method is suboptimal and relevant information is lost by integrating the signal trace. (3) We investigate how the departure from the true model parameters affects the fidelity of the HMM method.

We first review the PSB readout and the threshold method, then introduce the HMM method, and finally analyze the methods' fidelities.

\subsection{Pauli spin blockade readout}
Spin-qubit readout involves spin-to-charge conversion and measuring the charge with a charge sensor. In the PSB readout \cite{Ono2002, Hanson2007, Burkard2023, Oakes2023, Seedhouse2021},
the spin-to-charge conversion requires an ancilla quantum dot with a second electron spin. If the two spins form a singlet state, both electrons can be on the same quantum dot. If the two spins form a triplet state, tunneling of one electron to the quantum dot with the other electron is blocked due to the Pauli exclusion principle. This results in a different charge configuration than the singlet state. By measuring the charge configuration with a nearby charge sensor, one can distinguish triplet and singlet states and conclude about the initial spin. In this article, we investigate the \textit{data analysis} of the charge sensor signals \textit{after} the spin-to-charge conversion. Errors in this analysis are caused by noise in the charge sensor and by triplet-singlet relaxations during the measurement \cite{Takeda2024}. Throughout the article, we treat the qubit states as classical states; we assume that a projective measurement collapses the state whenever the charge sensor produces a signal. We also assume the signals are digital. 
This sets us apart from the extensive literature on weak continuous measurements \cite{vanHandel2005, Wiseman2009} 
and allows for a simple technical analysis using finite-size matrix algebra instead of It\^{o} calculus \cite{Ng2014}.

\newcommand{\ySum}{\overline{y}}

\newcommand{\yThreshold}[1]{#1_0}

\newcommand{\integrationTime}{\overline{T}}

\subsection{Threshold method}
The charge sensor circuitry emits a signal $y_t$ (a real number) at regular intervals indexed by integer $t$ with time step $\Delta$ apart set by the sampling rate. The vector of these discrete signals of length $T$ is denoted by $\{ y_t \}_{t=0}^{T-1}$ and here called the \textit{signal trace}. The total single-measurement data-collection time is $T\Delta$.

The signal traces are often processed using the threshold method. For PSB, it means the first $\integrationTime$ signals of a signal trace are averaged (equivalently, summed) producing the integrated signal $\ySum$. The qubit is read out as the triplet or singlet according to whether $\ySum$ is above or below a threshold value $\yThreshold{\ySum}$ fixed in advance \cite{Takeda2024, Oakes2023}.
Both $\yThreshold{\ySum}$ and $\integrationTime$ need to be optimized before the method is used for actual qubit-state assignments. The optimization of $\integrationTime$ balances two requirements. The integration time should be sufficiently long to reduce the effect of noise through averaging, and should be sufficiently short to reduce the number of triplet relaxations during the measurement. %

\begin{figure}
\includegraphics[width=0.45\textwidth]{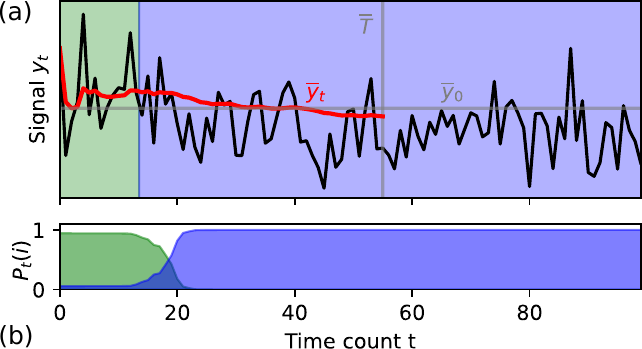}
\caption{(a) A simulated signal trace for the PSB readout. The initial state is a triplet (green). Between times $t=13$ and $t=14$, the triplet decays into a singlet state (blue). The threshold method gives a false state assignment, because integrating the signals (red) up to $\integrationTime$ 
results in a subthreshold value $\ySum<\yThreshold{\ySum}$. (b) In contrast to the threshold method, the HMM method delivers the state probabilities $P_t(i)$ for every $t$ and discloses the state transition. The signal trace is from the test data set used in Fig.~\ref{fig:mouse_fidelity_vs_A} with $A_{12}=0.0022$ and $\mathrm{SNR}=1$. We show only the first $100$ time points and define $\overline{y}_t=\frac{1}{t+1}\sum_{t'}y_{t'}$.}
\label{fig:signal_example_PSBreadout}
\end{figure}

By integrating the signal trace, information is lost because (i) many signal values are compressed into one new one and (ii) the signals after the integration time $\integrationTime$ are ignored. Retaining the complete information content can improve the readout fidelity.
A typical example of a signal trace is shown in Fig.~\ref{fig:signal_example_PSBreadout}(a), with the qubit relaxing early within the integration window. While the relaxation event is visible in the time dependence of the signal trace, a simple average results in a false subthreshold value and wrong state assignment.

\subsection{The hidden Markov model}\label{sec:c2}

A hidden Markov model \cite{Rabiner1989,Ephraim2002,Baum1970, Baum1972,NumericalRecipes} is a natural description of the readout described above, adding to it a simple qubit dynamics. We assume that at each value of the integer time index $t$, the qubit can be assigned a definite state $s_t$. There are $M$ available states, and we enumerate them by index $i\in \{1, 2, \ldots M\}$. The dynamics is described by the probabilities $A_{ij}$ for a transition from state $i$ to $j$ that can occur during the time interval $\Delta$ on advancing the time index by one step. The probability that the system is initially in state $i$ is denoted by $\pi_i$. The states are not directly observable and have to be inferred from observing the signals $y_t$. These are also emitted at times $t$ depending on the system state $s_t$: The probability that the system sends out the signal $y$ given that it is in state $i$ is denoted by $b_i(y)$. We assume it is a Gaussian function, 
\begin{align}
\label{eq:b(y)}
b_i(y)=(2\pi\sigma_i^2 )^{-1/2}\exp[-(y-\mu_i)^2/(2\sigma_i^2 )],
\end{align}
with the state-dependent mean $\mu_i$ and variance $\sigma_i^2$. This is not a principal restriction, as a straightforward extension to HMMs with mixtures of Gaussian functions can approximate other functional forms for $b(y)$ \cite{Juang1985,Rabiner1989,Ephraim2002}.
The parameters $A_{ij}$, $\mu_i$, $\sigma_i^2$, and $\pi_i$ are the HMM parameters, summarized as $\lambda=(\pi_i, \mu_i, \sigma_i^2, A_{ij})$. 

We describe the PSB readout as an HMM with two states, triplet ($i=1$) and singlet ($i=2$) with corresponding signals, $\mu_1=1$ and $\mu_2=0$. Transitions occur with probabilities $A_{12}>0$ and $A_{21}= 0 $ per time step. Noise broadens the signal according to variances $\sigma_i^2$. For simplicity, we take $\sigma_1^2 =\sigma_2^2\equiv\sigma^2 $, which allows us to define a signal-to-noise ratio $\mathrm{SNR} = |\mu_1-\mu_2|/\sqrt{\sigma^2}$.

With the above description, one can use an HMM to simulate readout, producing samples of signal traces: Choose the initial state according to $\pi_i$, evolve it stochastically for $T$ time steps according to the elements of matrix $A$, and assign a stochastic signal to each $s_t$ according to Eq.~\eqref{eq:b(y)}. Below, we use this procedure to generate measurement signals with white noise. The procedure can thus be viewed as assigning (one possible sample of) a signal trace to a given vector of system states. 

Now we describe an algorithm that aims at the inverse: given the vector of observations, infer the hidden states. The so-called forward-backward algorithm \cite{Rabiner1989,Baum1970, Baum1972} evaluates the exact conditional probability \cite{JaynesProbabilityTheory} that the system is in state $s_t=i$ at time $t$ given all available observations:
\begin{align}
\label{eq:P}
P_t(i)\equiv P(s_t=i|y_0,y_1,...,y_{T-1}, \lambda)= \frac{\alpha_t(i)\beta_t(i)}{\mathcal{L}}.
\end{align}
Here, $\mathcal{L}$ is the likelihood of the data, which describes how likely the observations are, given the model parameters:
\begin{align}
\label{eq:L}
\mathcal{L} = P(y_0, y_1,...,y_{T-1}|\lambda) = \sum_{i=1}^{M} \alpha_t(i)\beta_t(i)\qquad \forall t.
\end{align}
The forward variable $\alpha_t(i)  = P(y_0,...,y_t, s_t=i|\lambda)$ is the probability of observing the signals up to time $t$ and having the system in state $i$ at time $t$ given the model parameters. It can be calculated in a forward-recursive way:
\begin{align}
\label{eq:alpha}
\alpha_0(i) = b_i(y_0)\pi_i,\qquad \alpha_{t+1}(i)=\sum_{j=1}^{M}\alpha_t(j)A_{ji}b_i(y_{t+1}).
\end{align}
The backward variable $\beta_t(i)  = P(y_{t+1},...,y_{T-1}| s_t=i,\lambda)$ is the probability of observing the signals after time $t$ given that the system is in state $i$ at time $t$ and given the model parameters. It can be calculated in a backward-recursive way:
\begin{align}
\label{eq:beta}
\beta_{T-1}(i)&=1,\qquad \beta_{t-1}(i)=\sum_{j=1}^{M}A_{ij}b_j(y_t)\beta_t(j).
\end{align}
The state probabilities for the example signal trace in Fig.~\ref{fig:signal_example_PSBreadout}(a) are shown in Fig.~\ref{fig:signal_example_PSBreadout}(b) for reference.

To read out the qubit, we are interested in the state probabilities $P_{t=0}(i)$ at the beginning of the measurement, given all the signals that follow at $t\geq0$. If $P_{t=0}(i=1)>P_{t=0}(i=2)$, the qubit is read out as a triplet, otherwise the qubit is read out as a singlet.  
If the signals were indeed sampled from an HMM, the resulting readout fidelity will be optimal. However, we can also naively apply the forward-backward algorithm to signal traces that do not originate from an HMM, assuming that they resemble them. 
In that case, the performance of the forward-backward algorithm needs to be tested numerically or experimentally. We call the method of using the forward-backward algorithm \footnote{To calculate the initial state probabilities, actually we only need the backward variable, not the forward variable. However, the term ``backward algorithm'' is not used so frequently, therefore we keep the name ``forward-backward algorithm''. The forward variables are needed for the state probabilities with $t>0$ in Fig.~\ref{fig:signal_example_PSBreadout}(b) and for the Baum-Welch algorithm.} to assign the initial qubit states the HMM analysis or the HMM method, regardless of the origin of the signals.

There are open-source libraries that implement common HMM algorithms. We use the Python package \mbox{\footnotesize{HMMLEARN}} \cite{hmmlearn}.  As one can see from Eqs.~\eqref{eq:P}--\eqref{eq:beta}, the evaluation of the state probabilities $P_t(i)$ scales with $\mathcal{O}(TM^2)$. Concerning the scaling with $T$, which is a large number, both methods scale linearly. The computational overhead of the HMM versus threshold is thus mild, and for small $M$ might turn out to be inconsequential for practical applications.

 \subsection{ Parameter calibration} \label{sec:BW}

\newcommand{\bw}[1]{#1^*}

\begin{figure}
\includegraphics[width=0.45\textwidth]{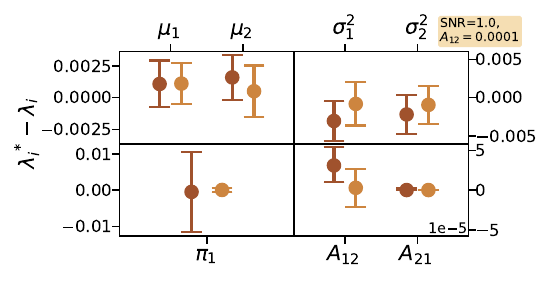}
\caption{Parameter residuals of one exemplary set of parameters in Fig.~\ref{fig:mouse_fidelity_vs_A}(a), calibrated by the Baum-Welch algorithm. We obtain confidence intervals of these parameters via the Monte Carlo (light brown) and the likelihood ratio (dark brown) methods. The parameters marked in dark brown were estimated using 2000 signal traces and were used by the HMM$^*$ fidelity calculations in Fig.~\ref{fig:mouse_fidelity_vs_A}. For the Monte Carlo confidence intervals, we used $5\times2000$ signal traces. }
\label{fig:PSB_example_residuals}
\end{figure}

Both the threshold method and the HMM method need calibration before they can be applied to readout data. By calibration we mean setting values for the parameters of the method including the statistical model it contains. Unlike in computer simulations, in experiments, even if the model is a fully adequate description of the system, its parameters are \textit{a priori} unknown and have to be estimated. This problem should not be ignored in the testing and comparison of different readout processing methods.

In experiments, the threshold method is usually calibrated by fitting the histograms of observed integrated readout signals by Gaussians, and optimizing  $\integrationTime$ so they have minimal overlap. The threshold value  $\yThreshold{\ySum}$ is given by the intersection of the Gaussians. In our simulations, we can avoid the additional uncertainty from such fits since we know the true initial states. We thus calibrate the parameters by maximizing the fidelity of assigning the correct initial state over the readout traces from a \textit{training set} (we use training sets of the same size as the \textit{test set}s described below).

HMM does not contain any parameters except those of the statistical model $\lambda$. The values for the latter can be assigned as those most compatible with the observed samples of readout traces. Specifically, one maximizes the likelihood of the training set as a function of the model parameters \footnote{As a side remark, we note that this procedure implies that the parameters' priors are ignored.}. Being a high-dimensional maximization, this is in general a complicated problem. Another appealing property of the HMM model is the existence of a robust general algorithm for this task. It is called Baum-Welch algorithm \cite{NumericalRecipes,Rabiner1989, Ephraim2002,Juang1985, House2009} and is a special case of the estimation-maximization algorithm \cite{Dempster1977}. It uses a training set as the input, and proceeds iteratively, increasing the likelihood at every iteration towards a local maximum, which often turns out to be the global maximum. The details of this algorithm are reviewed in Appendix~\ref{sec:baum_welch_algorithm}. 
We denote the parameters obtained from the converged Baum-Welch algorithm with an asterisk, $\bw{\lambda}=\{\bw{A}_{ij},\, \bw{\mu}_i,\,{\bw{{\sigma^2_i}}},\,\bw{\pi}_i\}$. Similar to the optimization of the threshold parameters, we run the Baum-Welch algorithm on training signals to avoid overfitting, using $2000$ signal traces.
In addition, we calculate confidence intervals of the model parameters using two methods, one based on the log likelihood curvature at the maximum (likelihood ratio confidence intervals) and one based on repeated sampling (Monte Carlo confidence intervals). They are explained and compared in Appendix~\ref{sec:baum_welch_algorithm}.

We find that the seven HMM parameters of PSB readout are estimated precisely by the Baum-Welch algorithm. An exemplary set of estimated parameters is shown in Fig.~\ref{fig:PSB_example_residuals}, including confidence intervals. 
Among the estimated parameters is the excitation probability $A_{21}$, here set to zero. This result emphasizes the flexibility of the HMM to describe systems where both relaxation and excitation rates can strongly vary in magnitude.

\subsection{Readout fidelity}

\begin{figure}
(a)\includegraphics[width=0.45\textwidth]{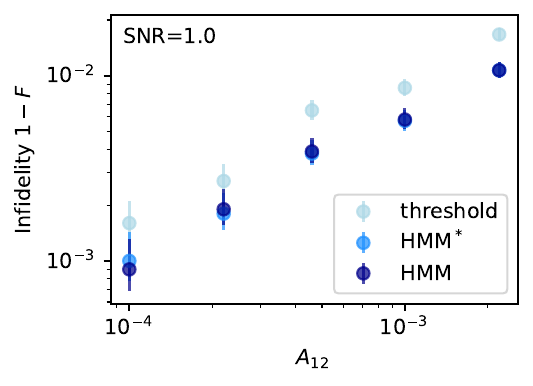}
(b)\includegraphics[width=0.45\textwidth]{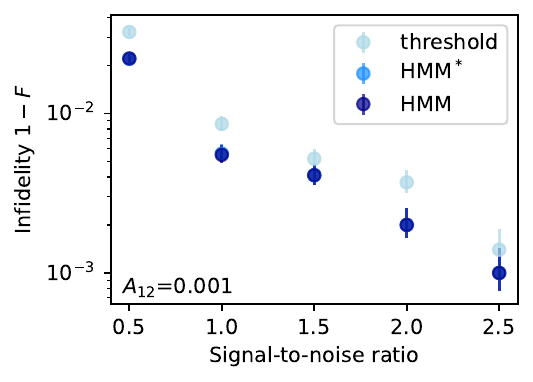}
\caption{Infidelity for PSB readout as a function of (a) relaxation probability per time step $A_{12}$ and (b) signal-to-noise ratio, calculated using simulated signals with white noise.
The points show the threshold method (light blue); HMM$^*$ (darker blue) stands for an HMM analysis using the model parameters estimated by the Baum-Welch algorithm, and HMM (darkest blue) for using the exact model parameters. }
\label{fig:mouse_fidelity_vs_A}
\end{figure}

After these preliminaries, we are in a position to evaluate the readout fidelity $F$ of the two introduced methods. To this end, we generate computer-simulated readout signal samples, as described in Sec.~\ref{sec:c2}. By a \textit{test set}, we mean a collection of a large number $N$ of traces ($N=10000$ unless noted otherwise), half of them initialized as singlet states, the other half as triplet states. The signal traces' length $T$ is chosen sufficiently long to not be a limitation for the fidelity ($T=300$ unless noted otherwise). Applying separately the threshold and HMM method over the test set, we infer the initial state from each signal trace and assign the infidelity ($1-F$) as the proportion of false assignments.
For very small infidelities, the fact that $N$ is finite implies a sizable statistical uncertainty on the estimated infidelity. We display the corresponding error bars in the figures \footnote{
We calculate this uncertainty numerically as follows. Assuming the method has fidelity $F$ and infidelity $p=1-F$, the number of false spin assignments $n$ follows a binomial distribution $P(n|p,N)=\begin{psmallmatrix}N\\ n\end{psmallmatrix}p^{n}(1-p)^{N-n}$. To estimate $p$ we need $P(p|n,N)= \frac{P(n|p,N)P(p|N)}{\int \text{d}p \, P(n|p,N)P(p|N)}$. Assuming a constant prior $P(p|N)$ which cancels out, $P(p|n,N)$ has a maximum at $\bw{p}=n/k$ and we numerically calculate a confidence interval $[a,b]$ for $p$ corresponding to a  $1-\alpha=68\%$ confidence level such that  $\int_0^aP(p|nN)\text{d}p=\alpha/2$  and $\int_b^1P(p|nN)\text{d}p=\alpha/2$.
}.

We compare the two methods in Fig.~\ref{fig:mouse_fidelity_vs_A}. The HMM has higher fidelity. This is not a coincidence since the forward-backward algorithm delivers the exact probability conditioned on the observed data. No processing can reach higher fidelity, \textit{assuming that the statistical model with which the data were produced and with which they were processed by the HMM are identical}. The threshold method does not reach the maximal possible fidelity, in agreement with Refs.~\cite{Gambetta2007, DAnjou2014}. As already explained, it loses information by omitting the time order of the signals. 
We observe that the gap (on the logarithmically scaled $y$ axis) between the threshold and HMM methods' infidelities remains similar throughout the investigated ranges, including small relaxation rates [Fig.~\ref{fig:mouse_fidelity_vs_A}(a)] and high SNRs [Fig.~\ref{fig:mouse_fidelity_vs_A}(b)].
This is both encouraging and surprising, since at infinite SNR \footnote{At infinite SNR, the fidelty of both methods is one.} or at strictly zero transition rate \footnote{Without transitions, the integrated signal is a sufficient statistic. We show this in  App.~\ref{sec:optimal_threshold}.}, the threshold method is optimal and there can be no gain from the HMM analysis.

We include fidelities assigned using the HMM with the estimated parameters $\bw{\lambda}$, labeled with an asterisk in the figure legends. The infidelities overlap with the infidelities calculated with the correct parameters, up to statistical fluctuations. It confirms that for white-noise signals, calibration through the Baum-Welch algorithm delivers essentially exact parameters, and does not degrade the method's fidelity.

\section{Time-correlated noise  }\label{sec:correlated_noise}

In the previous section, we have seen that for white noise, the HMM significantly improves the readout fidelity compared to the threshold method. However, the noise in real experiments is correlated. We now investigate how the forward-backward algorithm performs in this case. We again simulate PSB readout as a prototypical scenario.

\subsection{Noise spectrum}

To describe and later generate correlated noise, let us first assume we only have one state and no transitions, in which case we can follow Ref.~\cite{GutierrezRubio2022}. The autocorrelation function is defined as 
\begin{align}\label{eq:Sigma_ttprime}
\Sigma_{tt'}=\langle y_t y_{t'}\rangle - \langle y_t\rangle\langle y_{t'}\rangle.
\end{align}
Here the expectation value denotes the statistical average over the ensemble of signal traces. We first assume stationary noise, which gives that $\langle y_t\rangle=\mu $ is the same for all time steps $t$, and $\Sigma_{tt'}\equiv\Sigma_{t-t'}$ only depends on $t-t'$. We also make the approximation that the noise spectral density is negligible below the frequency $1/T$, so we can impose periodic boundary conditions $y_{t}=y_{t+T}$ for each trace. The Fourier transform of the autocorrelation function, called spectrum, then becomes \cite{GutierrezRubio2022} 
\begin{align}\label{eq:Lambda_k}
\Lambda_k \equiv \sum_{t=0}^{T-1}\Sigma_t \exp(-i2\pi t k /T), 
\end{align}
and is real and periodic, $\Lambda_k= \Lambda_{T-k}$. We generate signals with correlated noise from a given spectrum $\Lambda_k$ and mean $\mu$ as explained in Ref.~\cite{GutierrezRubio2022} and reviewed in Appendix~\ref{sec:generate_correlated_noise}. 

To analyze a concrete example, we choose a Gaussian-shaped spectrum,
\begin{align}
\Lambda_k &= \Sigma_0 T_c \sqrt{\pi}\exp[-(k \pi T_c / T)^2 ] \quad\text{for } 0\leq k\leq\lfloor{T/2}\rfloor,\nonumber \\
\Lambda_{k} &= \Lambda_{T-k} \qquad\qquad\qquad\qquad\quad\quad\text{else,  }
\label{eq:Lambda}
\end{align}
where $\Sigma_0$ is the sample variance and we introduced a parameter $T_c$. We interpret $T_c$ as the correlation time (in units of $\Delta$) because for $1\lesssim T_c\ll T$ the corresponding autocorrelation function is approximately $\Sigma_j\approx\Sigma_0\exp[-(j/T_c)^2 ]$ for $j\leq\lfloor T/2\rfloor$. We define the SNR as $|\mu_1-\mu_2|/\sqrt{\Sigma_0}$. In Appendix~\ref{sec:generate_correlated_noise}, we show examples of the spectrum and the autocorrelation function for a few values of $T_c$. 


The actual readout involves several states (two in PSB). To allow for, at least in principle, individual noise spectra for each state, we first generate signal traces $y_t^{(i)}$, $0\leq t\leq T-1$ for each of the states $i$. Next, we sample a state sequence $s_t$ based on the transition probabilities $A_{ij}$ and a chosen initial state $s_0$. The final signal trace $y_t=y_t^{(i=s_t)}$ is then composed of stretches of $y_t^{(i)}$. 

\subsection{Readout fidelity with correlated noise}

\begin{figure}
(a)\includegraphics[width=0.45\textwidth]{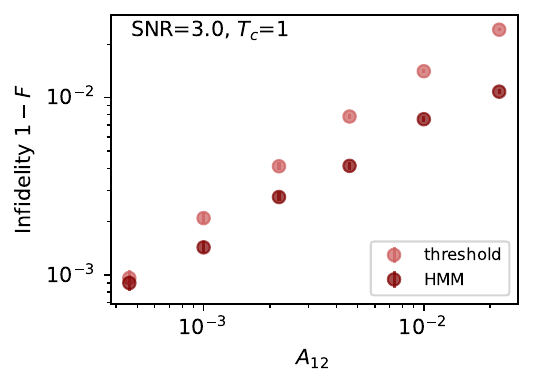}
(b)\includegraphics[width=0.45\textwidth]{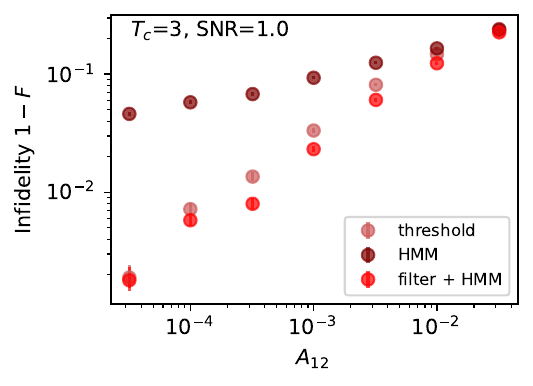}
\caption{Infidelity as a function of the relaxation probability for simulated signals with time-correlated noise. 
(a) The HMM and threshold methods for noise correlation time $T_c=1$. The plotted data were obtained from $100000$ signal traces with $T=30$ time steps. (b) The standard HMM, the HMM on preprocessed data by an averaging filter with $t_s=20$, and the threshold methods for noise correlation time $T_c=3$.
The plotted data were obtained from $10000$ signal traces with $T=300$ time steps. The effective model-matched SNR after the signal filter is SNR $=2.02$. The optimized analysis times of the threshold method are 283, 191, 192, 125, 86, 51, and 24 time steps for the increasing values of $A_{12}$ shown.
}
\label{fig:salamander_fidelity_corrlen1}
\end{figure}

Although the signals generated using correlated noise do not fulfill the HMM assumptions, we can still process the signal traces by it. We first examine the model calibration and then its performance in terms of fidelity. Concerning the former, there are no ``exact'' parameters to uncover since the statistical model inside the HMM is different from the one used to generate the data. Nevertheless, we consider the following HMM ``model-matched'' parameters to represent the highest possible correspondence (within the limits due to the model mismatch): (i) the initial state probabilities $\pi_i$, the transition probabilities $A_{ij}$, and the signal means $\mu_i$ are the same as in the correlated-noise signal, and (ii) the signal variances of the white noise of the HMM $\sigma_i^2$ are equal to the signal variance $\Sigma_0$.

First, we investigate if the Baum-Welch algorithm identifies the model-matched parameters. We find that it does not. For the noise correlations parameterized by Eq.~\eqref{eq:Lambda}, even a correlation time of $T_c=1$ is fatal for the calibration: For some parameters, the relative deviation between their model-matched value and the one estimated by the Baum-Welch algorithm is of the order of $1$. The quantitative details are given in Appendix~\ref{sec:correlation_baumwelch}; see Fig.~\ref{fig:parameter_estimation_correlation}. At longer correlation times, the discrepancy grows further. 

We thus give up on a self-consistent calibration and assume that the model-matched parameters are known from some separate experimental calibration procedure. Under this optimistic scenario, we compare the HMM and threshold method fidelities in assigning the state. We find that the HMM is better at small correlation times (of the order of a single time step); see  [Fig.~\ref{fig:salamander_fidelity_corrlen1} (a)]. Also, large transition rates and  SNRs increase the performance of the HMM relative to the threshold method. The reason is, again, that the threshold method does not reflect the state transitions correctly. In addition, at high SNR the HMM only requires a few time steps to determine the state and thus detects transitions during the readout more accurately.

On the other hand, the HMM performance becomes drastically worse as the correlation time increases [Fig.~\ref{fig:salamander_fidelity_corrlen1} b)]. The presence of correlations makes the forward-backward algorithm overestimate the number of transitions, as it misinterprets correlations as transitions. The inability of the HMM to keep up with the threshold method is particularly pronounced at small SNRs and large relaxation probabilities $A_{12}$. 
This important result shows that analytical as well as computer-simulated results based on white-noise models might have limited relevance to assess the performance of those algorithms on real-world data.

By Eq.~\eqref{eq:Lambda}, we have chosen noise with a limited extent of correlations; the noise autocorrelation drops superexponentially for time delays beyond $T_c$. In this case, one can hope to efficiently remove correlations by averaging. We implement a simple averaging filter, replacing $t_s$ consecutive time points by their average, and examine the HMM performance on such prefiltered data \footnote{This simplest imaginable filtering can be viewed as intermediate between two limit cases: no filtering on the one hand, and compressing the whole trace to a single averaged point on the other, the latter being essentially the threshold method.}.
The averaging shortens the signal trace to $\lfloor T/t_s\rfloor$ time points and requires redefining model-matched parameters. While the signal means $\mu_i$ and initial state probabilities $\pi_i$ remain equal to the corresponding ones of the original signal trace, the transition probabilities $A_{ij}$ increase by the factor of $t_s$ (assuming $A_{ij} t_s \ll 1$) in the filtered signals compared to the original signals. As for the model-matched signal variances $\sigma_i^2$, we calculate them numerically as sample variances of filtered signals with correlated noise generated without transitions, $A_{12}=A_{21}=0$.

The fidelity calculations with filtered signals are summarized in Fig.~\ref{fig:salamander_fidelity_corrlen1}(b). The filtering improves the HMM performance remarkably. For the given parameters and the figure range, the HMM method again surpasses the threshold method. However, we point out that the simple filter has limitations. First, once the filtering time $t_s$ required for the HMM method becomes longer than the threshold-method optimal time $\integrationTime$, the HMM method can not be better than the threshold. Second, and more importantly, it remains unclear how much the filtering helps for noise with different long-time correlation behavior (for example, for $1/f$ noise with a power-law decay of the autocorrelation function). We leave it as an open question and conclude that it is desirable to develop reliable optimized algorithms for readout with correlated noise.

\section{Elzerman readout with white noise}\label{sec:Elzerman_fidelity}

In the previous section, we have shown that applying HMM with a white-noise statistical model to data with correlated noise might yield qualitatively wrong results. 
Here we give an example where the HMM analysis looks appropriate. We consider readout at finite temperatures, where there are potentially many transitions during the readout time once the temperature reaches the Zeeman-energy scale. Linear and nonlinear optimal filters developed for the zero-temperature regime \cite{Gambetta2007} are then not applicable. We consider the Elzerman readout \cite{Elzerman2004}, %
for which the temperature effects can easily become appreciable.

In this section, we only consider white noise in the simulations because we assume that either the noise correlation in the signal is sufficiently small or that the signal can be approximated by an equivalent white-noise model with effective parameters (signal to noise ratios, transition rates). For example, the signal may be converted to such a model by a filter, as suggested in Sec. \ref{sec:correlated_noise}.

The Elzerman readout differs from PSB in two ways. First, there are three states involved: the dot occupied by a spin-up electron, by a spin-down electron, and the empty dot. The two spin states induce the same signal at the charge sensor and only differ by their transition rates. Second, the threshold method uses the peak instead of the mean of the signal trace. 
We first consider zero temperature and demonstrate results analogous to the PSB scenario: The HMM method performs significantly better than the threshold, in line with Ref.~\cite{DAnjou2014}, and the fidelities remain unaffected when the model parameters are calibrated with the Baum-Welch algorithm. After that, we analyze the Elzerman readout at finite temperatures.

\subsection{Elzerman readout}

Elzerman readout \cite{Elzerman2004, Hanson2007, Burkard2023, Oakes2023, Mills2022, Volk2019} takes advantage of the spin dependence of the dot-lead tunnel rates for electrons. It originates in the Zeeman energy and arises as follows. Using gate potential, the electron states' energies are positioned such that the spin-up level is above and the spin-down state is below the chemical potential of a nearby lead. If the initial state is a spin-up electron in the quantum dot, it will tunnel out and shortly after be replaced by a spin-down electron from the lead. The intermediate empty-dot state is detected as a temporary change (a blip) in the charge-sensor signal. In contrast, a quantum-dot spin-down electron does not have the energy to tunnel out, and there is no blip in the signal trace.

\newcommand{\yPeak}{\hat{y}}

\newcommand{\waitingTime}{\hat{T}}

\begin{figure}
\includegraphics[width=0.45\textwidth]{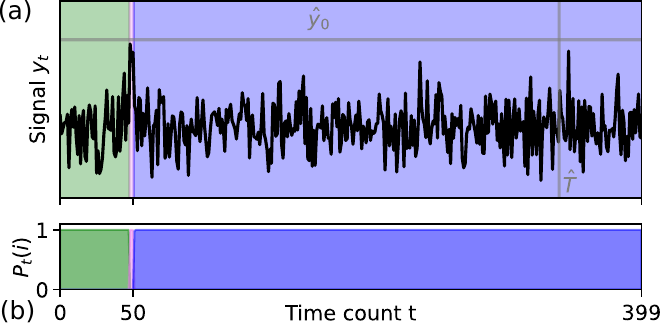}
\caption{ (a) Example of a simulated charge sensor signal trace for Elzerman readout. The initial state is a spin-up state (green) that transitions to an intermediate empty-dot state (rose) and a spin-down state (blue). The intermediate empty-dot state is very short-lived with unusually low signal values that do not exceed the threshold value $\yThreshold{\yPeak}$ within the waiting time $\waitingTime$. This causes a readout error when analyzed by the threshold method. (b) In contrast, the HMM analysis correctly identifies the initial spin-up state and the transition to the spin-down state.   
The signal trace is from the test data set used in Fig.~\ref{fig:fidelity_vs_SNR_A0.02} with SNR $=4$. }
\label{fig:signal_example_Elzermanreadout}
\end{figure}

\subsection{The threshold method for Elzerman readout}

The threshold method for Elzerman readout  \cite{Keith2019, Mills2022} detects the presence or absence of a blip based on comparing the signal-trace peak value $\yPeak$ to a threshold $\yThreshold{\yPeak}$ \cite{DAnjou2014}.
The value $\yPeak$ is the largest signal during the waiting time $\waitingTime$.
Both the threshold $\yThreshold{\yPeak}$ and the waiting time $\waitingTime$ need to be optimized. The calibration of the waiting time balances two requirements. It should be sufficiently short to avoid random fluctuations causing a false above-threshold signal, and it should be sufficiently long to leave time for a spin-up electron to tunnel out of the quantum dot.
Similar to PSB readout, the threshold method of Elzerman readout loses information by ignoring everything except the signal-trace maximum and all values after the waiting time. We show an example signal trace for which the threshold method fails in Fig.~\ref{fig:signal_example_Elzermanreadout}(a). The HMM model, which takes into account the full dynamics of the states, should improve the fidelity.

\subsection{HMM setup for Elzerman readout}

We describe the Elzerman readout as an HMM with three states: an occupied dot with spin up ($i=1$), an empty dot ($i=2$), and an occupied dot with spin down ($i=3$). The two occupied states cause the same charge-sensor signal, $\mu_1=\mu_3=0$, $\sigma_1^2 =\sigma_3^2$, and can only be distinguished by their transition rates modeled via $A_{12}, A_{23}\gg A_{21}, A_{32}$. The empty-dot state causes a larger charge-sensor signal, $\mu_2=1$. 
Direct spin flips can, in principle, be described by the HMM, but are typically small, $A_{13},\ A_{31}\approx 0$. For the zero temperature calculations, we choose $A_{12}=A_{23}\neq0$ and all other $A_{ij}=0$.  We set $\pi_1=\pi_3=0.5$ and do not estimate $\pi_i$ with the Baum-Welch algorithm because for this degenerate model several parameter sets with different $\pi_i$ exist that result in the same likelihood value. We generate signal traces with the length of $T=400$ time steps, unless noted otherwise, and show an example calculation of the forward-backward algorithm in Fig.~\ref{fig:signal_example_Elzermanreadout}(b).

\subsection{Elzerman readout fidelities at zero temperature}

\begin{figure}
    \includegraphics[width=0.45\textwidth]{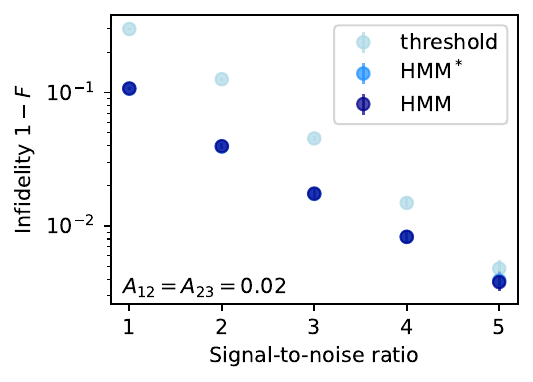}
    \caption{
    Infidelity for the Elzerman readout as a function of the SNR.
    The HMM method has smaller infidelity than the threshold method (light blue), even at large SNRs. The infidelities calculated using the parameters calibrated by the Baum-Welch algorithm (darker blue) overlap with those obtained using the exact model parameters (darkest blue). We used sets with $10000$ signal traces to estimate the methods' infidelities and to calibrate the threshold method parameters and we used sets with $2000$ signal traces to calibrate the HMM method. The calibrated parameters are shown in Figs. \ref{fig:mouse_residuen_A0_SNR2} and \ref{fig:mouse_residuen_A0_SNR0} (dark brown points).}
    \label{fig:fidelity_vs_SNR_A0.02}
\end{figure}

We present the infidelities for Elzerman readout at zero temperature for various SNRs in Fig.~\ref{fig:fidelity_vs_SNR_A0.02}.  The threshold method has significantly larger infidelity than HMM for SNRs up to 4. This demonstrates that the threshold method is not optimal and a significant amount of information is lost by only taking into account the peak value of the signal, in agreement with  Ref.~\cite{DAnjou2014}. 
At SNR $=5$, the difference between the methods appears to decrease, although the statistical error is too large to make a definite conclusion. At even larger SNRs, we expect that the infidelity will be limited by cases where the spin-up electron leaves the dot and within less than one time step a spin-down electron enters the dot. This process stays unnoticed by the charge sensor and we neglected it in our simulations by setting $A_{13}=0$.

\begin{figure}
\includegraphics[width=0.45\textwidth]{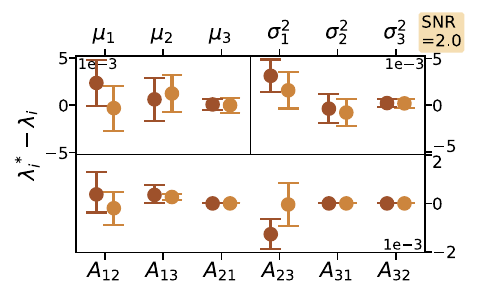}
\caption{Parameter residuals of an exemplary set of the Elzerman readout parameters in Fig.~\ref{fig:fidelity_vs_SNR_A0.02}. The confidence intervals were calculated with the likelihood ratio method (dark brown) and Monte Carlo method (light brown), analogous to the confidence-interval calculations for PSB readout.}
\label{fig:mouse_residuen_A0_SNR2}
\end{figure}

We also run the Baum-Welch algorithm and find that all 12 HMM model parameters are estimated successfully \footnote{Only for unrealistic large transition rates of $A_{12}=A_{23}=0.18$ and a small SNR $=0.5$ we observed cases where the Baum-Welch algorithm did not converge to the correct parameters.}. 
An exemplary set of parameters is shown in Fig.~\ref{fig:mouse_residuen_A0_SNR2}. The algorithm also correctly identifies the nontrivial feature that two states produce the same signals and only differ by their transition rates, as studied before in Ref.~\cite{House2013}. 
We repeat the HMM analysis using the estimated parameters instead of the exact ones, and find no relevant change in the resulting infidelities; see Fig.~\ref{fig:fidelity_vs_SNR_A0.02}.

\subsection{Elzerman readout at elevated temperatures}

Elzerman readout works best if only spin-up states can tunnel out of the QD, which is the case for low temperatures. Once the thermal energy becomes of the order of the Zeeman energy $E_Z=\frac{1}{2}g\mu_B B$ \footnote{Here $B$ is the magnetic field, $g$ is the g-factor and $\mu_B$ is the Bohr magneton.}, spin-down electrons may also tunnel out of the dot, and spin-up electrons may tunnel into an empty dot \cite{Johnson2022, Gorman2017}.
We model this situation with HMM parameters \cite{House2013, Keith2019}
\begin{align}
A_{12}=&A_{23}=[1-f(E_Z/(k_B\mathcal{T}))]A_0, \label{eq:thermal_relaxation}\\
A_{21}=&A_{32}=f(E_Z/(k_B\mathcal{T}))A_0. \label{eq:thermal_excitation}
\end{align}
Here, $f(x)=1/[1+\exp(x)]$ is the Fermi function, $\mathcal{T}$ is the temperature, and $k_B$ is the Boltzmann constant. The transition probability at low temperature is described by $A_0$ and is tunable by gates. All other model parameters remain as before.

\begin{figure}
(a)\includegraphics[width=0.45\textwidth]{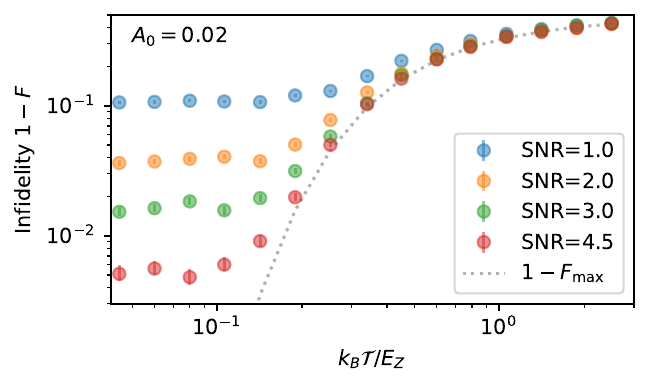}
(b)\includegraphics[width=0.45\textwidth]{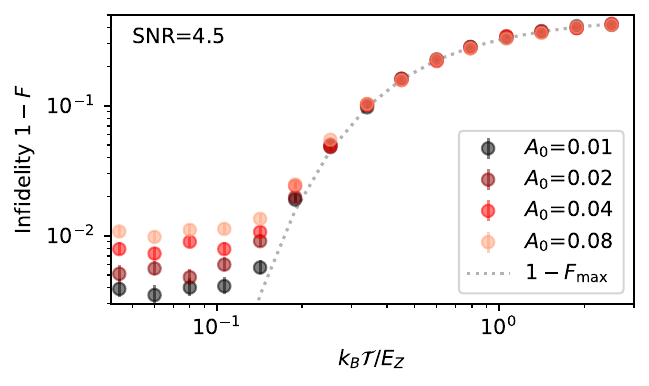}
\caption{Readout infidelities as a function of temperature for Elzerman readout.
The fidelity is limited by (a) the SNR and (b) the transition probabilities at small temperatures, and by temperature once the thermal energy becomes the Zeeman energy. In (b), we choose signal trace lengths $T$ of $800$, $400$, $250$, and $150$ for the increasing values of $A_0$. Both (a) and (b) use 10000 signal traces to evaluate the infidelity.
}
\label{fig:Firefly}
\end{figure}

We repeat the fidelity calculations, taking into account the effect of thermal excitations, and present the results in Fig.~\ref{fig:Firefly}. We see that at low temperature $k_B\mathcal{T}\ll E_Z$, the fidelity is limited by the SNR or the transition probabilities and hardly depends on the temperature. At high temperature $k_B\mathcal{T}\gtrsim E_Z$, the fidelity is limited by the temperature and hardly depends on the SNR or $A_0$. For example, for $k_B\mathcal{T} \gtrsim 0.4 E_Z$, the readout fidelity cannot be pushed above $\sim90\%$, irrespective of the SNR. 
The fidelities at high temperature are almost exclusively determined by the fact that the spin-up tunnel-out probability $A_{12}$ and the spin-down tunnel-out probability $A_{32}$ become more and more indistinguishable due to thermal excitations, set by Eqs. \eqref{eq:thermal_relaxation} and \eqref{eq:thermal_excitation}. We can model that dominant effect by taking the limit of zero noise. In that limit, the  charge detector detects any charge transition with certainty and the fidelity $F_\text{max}$ depends exclusively on the ratio $r=A_{12}/A_{32}$
\footnote{The result requires the reasonable assumption that transition probabilities per time steps fulfill $A_{12},A_{32}\ll1$ and is derived from $1-F_\mathrm{max}=\displaystyle\frac{1}{2}\sum_{t=0}^{\infty} \min_{A\in\{A_{12}A_{32}\}}(1-A)^tA$.}:
\begin{align}
1-F_\text{max} = \frac{1}{2}\left[1-r^{1/(1-r)}(1-r^{-1})\right].
\end{align}
At high temperature, the fidelities calculated with the HMM border on the theoretically-derived maximal fidelity $F_\text{max}$; see Fig.~\ref{fig:Firefly}.  Our HMM calculation visualize the transition between the high-temperature region and the low-temperature region.

\section{Conclusion}\label{sec:conclusion}

We have studied whether replacing the threshold method with an HMM analysis can improve the fidelity of spin-qubit readout. Our main finding is that the HMM based on a white-noise model is very sensitive to noise correlations and quickly degrades in their presence. Except for noise with very short correlation times, the performance of the HMM analysis drops significantly, particularly at small SNRs. This means that the benefits of the HMM processing \footnote{And also similar solutions provably optimal for the white-noise zero-temperature case, see, for example, Ref.~\onlinecite{Gambetta2007}.} might be illusory when applied to real-world data.  
One preliminary solution we have identified is to prepend the HMM processing by a simple moving-average filter, to reduce the effective correlation time of the noise. However, it is not clear to what extent such a strategy works for noises with long-range correlations, such as 1/$f$ noise. 

This drawback calls for working out HMM algorithms for correlated noise \cite{Almanjahie2014}.
We speculate that the increased computational complexity that time correlations introduce \cite{Hadar2009} makes solutions based on neural networks \cite{Struck2021} promising. %

We have also investigated the readout fidelities for the Elzerman readout at finite temperature. High temperature implies frequent state transitions during the readout, which justifies the use of an HMM analysis.  Upon increasing the temperature, the HMM maps out the transition from the fidelity being limited by a finite SNR to being limited by the temperature.

\textit{Note added.} Recently, we became aware of Ref. \cite{Laine2025}, where the authors used a HMM to distinguish two triplet and one singlet state during PSB readout and included a two-level fluctuator in their model to account for charge noise.

\subsection*{Acknowledgements}
This work was supported as a part of NCCR SPIN, a National Centre of Competence in Research, funded by the Swiss National Science Foundation (Grant No. 180604).

\appendix

\section{Special case of threshold method optimality} \label{sec:optimal_threshold}
Here we show that the threshold method is optimal for the following task. There are two states to differentiate, no transitions between them, and the two states correspond to signals with white Gaussian noise of the same variance. This result is helpful to understand when we can expect the fidelities for the threshold and HMM methods to converge. 

To this end, let us assume $T$ signals $y_t$, $0\leq t\leq T-1$, which either originate from state $A$ or state $B$ according to Eq.~\eqref{eq:b(y)}. The states are labeled $A$ and $B$ such that $\mu_A<\mu_B$. We further assume that both states are equally likely. The logarithm of the ratio of conditional probabilities for the two states, given the observations, is
\begin{align}
\ln R&\equiv\ln P(A|y_0,...,y_{T-1}) - \ln P(B|y_0,...,y_{T-1}) \nonumber\\
&= \ln \frac{P_A}{P_B}+T\ln \frac{\sigma_B}{\sigma_A} \nonumber\\
&\qquad\qquad\qquad +  \sum_{t=0}^{T-1} \frac{(y_t-\mu_B)^2}{2\sigma_B^2}-\frac{(y_t-\mu_A)^2}{2\sigma_A^2} \nonumber\\
&=  \ln \frac{P_A\sigma_B^T}{P_B\sigma_A^T}+ T \Bigg( \left[ \frac{\mu_B^2}{2\sigma_B^2}-\frac{\mu_A^2}{2\sigma_A^2}  \right] \nonumber\\
 &\qquad \qquad  + \overline{y^2} \left[ \frac{1}{2\sigma_B^2}-\frac{1}{2\sigma_A^2}  \right] + \overline{y} \left[ \frac{\mu_A}{\sigma_A^2} -\frac{\mu_B}{\sigma_B^2} \right]  \Bigg),
 \label{eq:logR}
\end{align}
where $P_A$ and $P_B$ are the prior probabilities of the two states, and we have defined the averages $\overline{y^n}=\frac{1}{T}\sum_{t=0}^{T-1} y_t^n$.

From the last line in Eq.~\eqref{eq:logR}, one can see that the exact conditional probability of the system state is a function of the signal mean and the signal variance. The two numbers, $\{\overline{y}, \overline{y^2}\}$, are thus the sufficient statistics. It is only in the case $\sigma_A=\sigma_B$ that the signal variance drops out and the sufficient statistic contains only the signal mean $\{ \overline{y} \}$. The more probable state is $A$ if $\ln R>0$ and $B$ if $\ln R<0$. The boundary between the two cases is given by $\ln R=0$, which for $\sigma_A=\sigma_B$ leads to a linear equation for the mean, $\overline{y} = \yThreshold{\ySum}$. We obtain that a threshold method is equivalent to examining the exact conditional probability, and is thus optimal (can not be improved further).

The deficiency of the threshold method for the case $\sigma_A\neq\sigma_B$ can be demonstrated with the following simple example. 
Let us assume $P_A=P_B=1/2$, $\mu_A=0$, $\sigma_A = 1.2$, $\mu_B=1$, and $\sigma_B=1$. In an unlikely but possible event that the signal trace consists of values $\vec{y}=\{10,0,0,0,0,0,0,0,0,0\}$, we find $\ln R>0$ because $P(y_t|A)>P(y_t|B)$  for all $t$, so state $A$ is more likely. In a different event, a signal trace may consist of values $\vec{y}=\{1,1,1,1,1,1,1,1,1,1\}$, clearly resulting in $\ln R<0$, thus state $B$ is more likely. However, in both events the signals have the same mean value $\overline{y}=1$. Thus, analysis methods that only take into account the integrated signal cannot output the exact conditional probability.

\section{Baum-Welch algorithm and confidence intervals} \label{sec:baum_welch_algorithm}

\subsection{Baum-Welch algorithm}
The Baum-Welch algorithm is an iterative algorithm to find the maximum likelihood parameters of an HMM \cite{NumericalRecipes,Rabiner1989, Ephraim2002,Juang1985, House2009}. The algorithm is robust, as the likelihood (of the current parameters' estimates) is guaranteed to increase with each iteration, until a maximum is found. Although the algorithm has been known for many decades, here we review it for completeness.

Let us assume our training set consists of $N$ signal traces \footnote{To obtain reasonable estimates of the parameters, the training set should consist of many signal traces. The reason is that some transitions or states may not occur a sufficient number of times or at all during a single signal trace.} with different states $s_{nt}$ and observations $y_{nt}$ where $1\leq n\leq N$. The state probabilities $P_{nt}(i)$, likelihood $\mathcal{L}_{n}$, forward variables $\alpha_{nt}(i)$ and backward variables $\beta_{nt}(i)$ are then calculated for each trace $n$, in the same way as for the forward-backward algorithm. The total likelihood of all signal traces  then is
\begin{align}
L=\prod_{n=1}^{N} \mathcal{L}_n.
\end{align}
The likelihood $L$ is usually an exponentially small number and therefore represented by its logarithm, LL$\,\equiv\ln(L)$. 

The Baum-Welch algorithm maximizes the total likelihood $L$. 
One iteration of the algorithm includes two steps:
First, one calculates the forward and backward variables $\alpha_{nt}(i)$ and $\beta_{nt}(i)$ and the state probabilities $P_{nt}(i)$ using the model parameters $\lambda$ from the previous iteration (or some initial choice when starting the algorithm \footnote{We initialize the Baum-Welch algorithm for PSB readout with $\pi_1=0.45$, $A_{12}^{init}=A_{21}^{init}=3\cdot 10^{-4}$, $\mu_1^{init}=0.4$, $\mu_2^{init}=0.3$, $\sigma_i^{2\ init}=0.36$. For Elzerman readout we initialize parameters at $A_{12}^{init}=A_{23}^{init}=2\cdot 10^{-4}$, $A_{13}^{init}=A_{21}^{init}=A_{31}^{init}=A_{32}^{init}=10^{-4}$, $\mu_1^{init}=\mu_3^{init}=0.3$, $\mu_2^{init}=0.4$, $\sigma_i^{2\ init}=0.36$. 
}).
Second, one reestimates the model parameters as
\begin{align}
\pi_{i,\text{new}}&= \frac{1}{N} \sum\nolimits_{n=1}^{N} P_{n1}(i),\\
\mu_{i,\text{new}}&=\frac{\sum_{n=1}^{N} \sum_{t=0}^{T-1}P_{nt}(i)\,y_{nt}}{\sum_{n=1}^{N} \sum_{t=0}^{T-1}P_{nt}(i)},\\
\sigma_{i,\text{new}}^2&=\frac{\sum_{n=1}^{N} \sum_{t=0}^{T-1}P_{nt}(i)\,y_{nt}^2}{\sum_{n=1}^{N} \sum_{t=0}^{T-1}P_{nt}(i)}-\mu_{i,\text{new}}^2,\\
A_{ij,\text{new}}&=\frac{\sum_{n=1}^{N} \mathcal{L}_n^{-1} \sum_{t=1}^{T-2}\alpha_{nt}(i)A_{ij}b_j(y_{n,t+1})\beta_{n,t+1}(j)}{\sum_{n=1}^{N} \mathcal{L}_n^{-1}\sum_{t=0}^{T-2}\alpha_{nt}(i)\beta_{nt}(i)  }.
\end{align}
We stop the Baum-Welch algorithm when the log likelihood (LL) increases less than 0.001 and mark the parameters obtained
with an asterisk, $\bw{\lambda}$.%

\begin{figure*}
\includegraphics[width=0.3\textwidth]{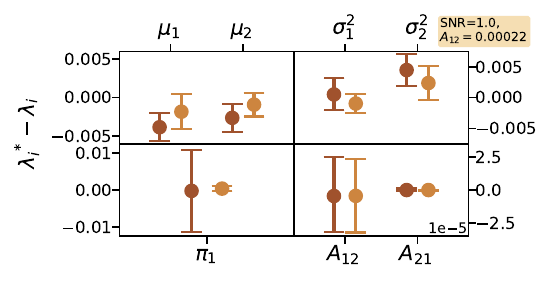}
\includegraphics[width=0.3\textwidth]{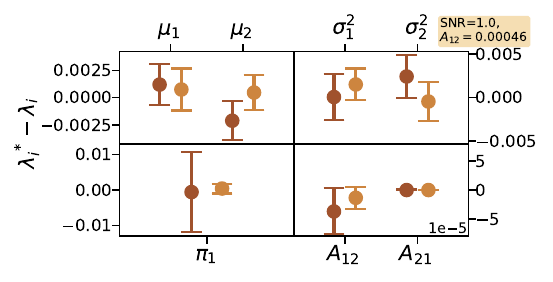}
\includegraphics[width=0.3\textwidth]{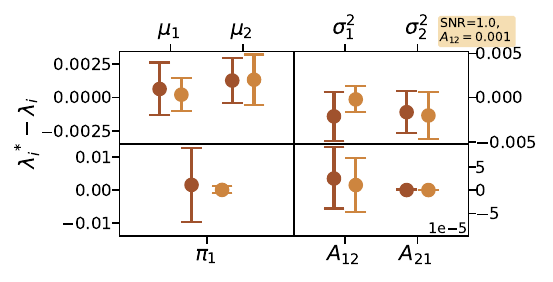}
\includegraphics[width=0.3\textwidth]{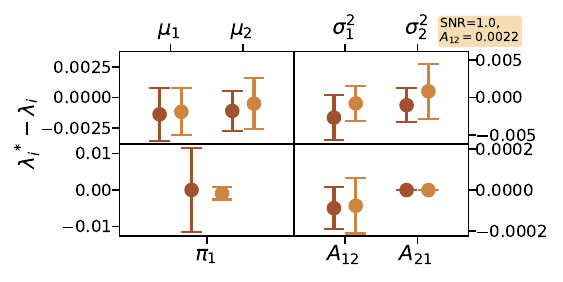}
\includegraphics[width=0.3\textwidth]{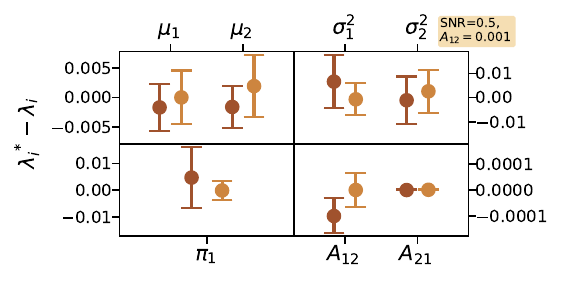}
\includegraphics[width=0.3\textwidth]{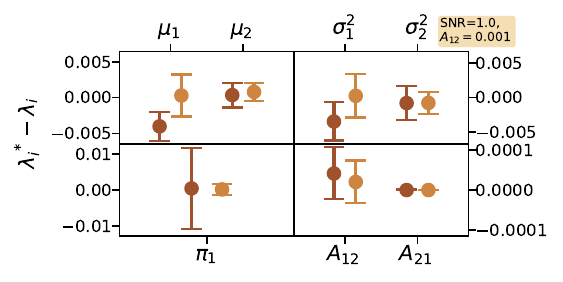}
\includegraphics[width=0.3\textwidth]{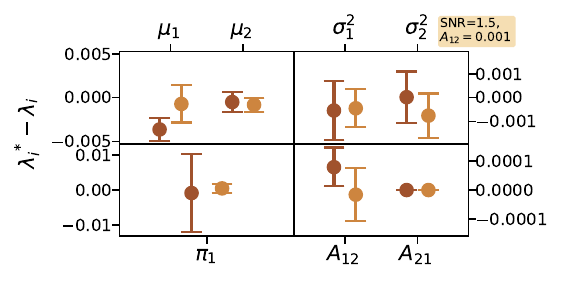}
\includegraphics[width=0.3\textwidth]{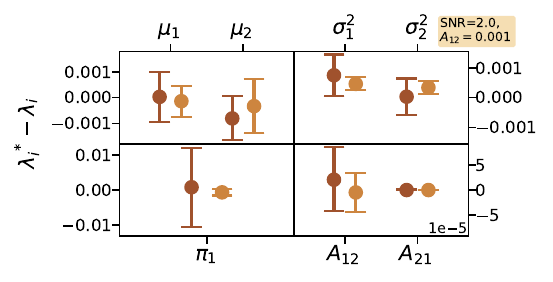}
\includegraphics[width=0.3\textwidth]{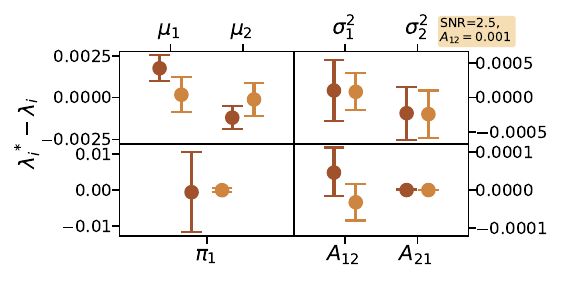}
\caption{Remaining parameter residuals  and confidence intervals of the Baum-Welch estimated parameters in Fig.~\ref{fig:mouse_fidelity_vs_A}, shown for completeness.}%
\label{fig:mouse_residuen_A1_SNR0}
\end{figure*}

\begin{figure*}
\includegraphics[width=0.24\textwidth]{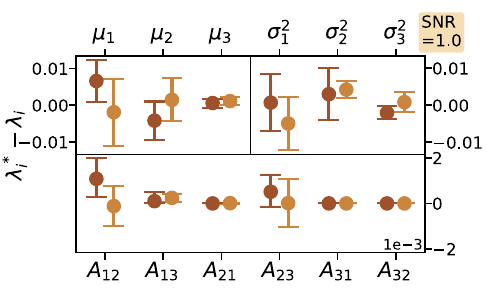}
\includegraphics[width=0.24\textwidth]{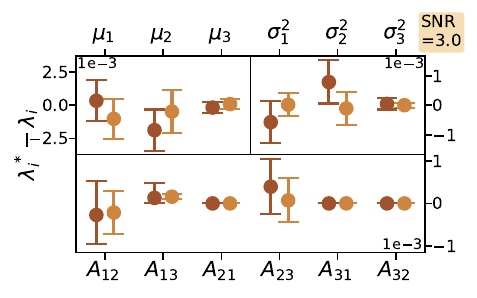}
\includegraphics[width=0.24\textwidth]{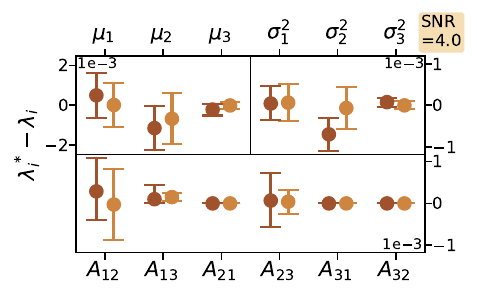}
\includegraphics[width=0.24\textwidth]{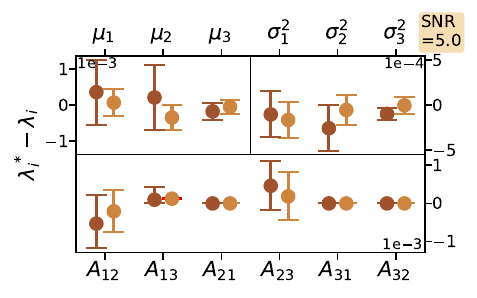}
\caption{Remaining parameter residuals and confidence intervals of the Baum-Welch estimated parameters in Fig.~\ref{fig:fidelity_vs_SNR_A0.02}, shown for completeness. 
}
\label{fig:mouse_residuen_A0_SNR0}
\end{figure*}

\subsection{Likelihood ratio confidence intervals}
While the likelihood maximum gives us the most likely model parameters $\bw{\lambda}$, its shape around that maximum describes how stable the estimate is with respect to statistical fluctuations in the training set. Integrating the likelihood function over a volume of the parameters' space provides the confidence level that the model parameters are in this volume. However, for a complicated multivariate LL function as we have for the HMM this integration is unfeasible. By assuming that the likelihood function approximately has a Gaussian shape around the maximum, one can simplify the problem, a method that is well established in statistics \cite{House2013, NumericalRecipes}. 
We follow this approach and evaluate the confidence interval of a model parameter $\lambda_0$ numerically, finding $\delta_L$ such that
\begin{align} \label{eq:logLratio_eq}
\max\limits_{\lambda_i}[\text{LL}(\lambda_i)] - \max\limits_{\lambda_i\neq\lambda_0}[\text{LL}(\bw{\lambda}_0+\delta_L,\lambda_1, \lambda_2,...)] = \Delta.
\end{align}
We choose a confidence level of $68$\%, which corresponds to a value of $\Delta=1/2$. Usually, there is a positive value, $\delta_L^+$, and a negative value, $-\delta_L^-$, that fulfill this equation and the resulting confidence interval is ${[\bw{\lambda}_0-\delta_L^{-}, \bw{\lambda}_0+\delta_L^{+}]}$ \footnote{We imposed a minimum value $\delta_L^{\pm}\geq3.4\cdot 10^{-7}$ because, given the amount of training data, we do not expect any algorithm to provide an accuracy better than that.}. Sometimes, $\delta_L$ cannot be found because it renders the parameter outside of its definition region, in which case we take the value at the region boundary.

\subsection{Monte Carlo confidence intervals}

A second way to find out how reliably the model parameters have been estimated is to repeat the estimation on several training sets and see how much these estimates differ from each other \cite{IntroductionStatisticalLearning}. We call this method the Monte Carlo (MC) method. If we have $N_d$ different data sets and estimate the parameter $\lambda_0$ as $\bw{\lambda}_{0,d}$ on data set $d$, then our confidence interval for this parameter is $[\bar{\lambda}_0-\delta_\text{MC},\bar{\lambda}_0+\delta_\text{MC}]$ where $\bar{\lambda}_0$ is the mean and $\delta_\text{MC}$ is the unbiased estimate of the standard deviation over those $N_d$ data sets \footnote{We again imposed a minimum value of $\delta_\text{MC}\geq3.4\cdot 10^{-7}$.}. Further, we take $\bar{\lambda}_0$ as the estimate for the parameter $\lambda_0$, which, as a mean value, technically is not a maximum likelihood estimate itself.%

\subsection{Comparison of confidence intervals calculations}

\begin{figure*}
(a)\includegraphics[width=0.4\textwidth]{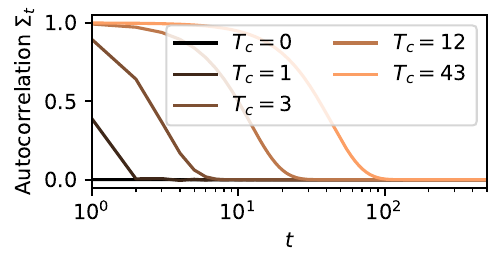}
(b)\includegraphics[width=0.4\textwidth]{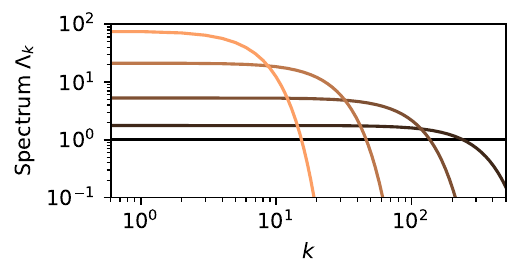}
\caption{(a) Autocorrelation functions $\Sigma_t$ used to generate signals with correlated noise, and (b) their Fourier transform, the spectrum $\Lambda_k$.  The correlation time $T_c$ indicates over how many time steps the data is correlated, with $T_c=0$ corresponding to uncorrelated (that is, white) noise. We choose $\Sigma_0=1$ and $T=1000$ and only show the independent components of $\Sigma_t$ and $\Lambda_k$, thus $t,k\leq\lfloor T/2\rfloor$. 
}
\label{fig:gaussianfrequency_correlation}
\end{figure*}

In the presence of white noise, the Baum-Welch estimated parameters and their confidence intervals are overall reliable, as shown for PSB readout in Figs.~\ref{fig:PSB_example_residuals} and \ref{fig:mouse_residuen_A1_SNR0} and for Elzerman readout in Figs.~\ref{fig:mouse_residuen_A0_SNR2} and \ref{fig:mouse_residuen_A0_SNR0}. The true parameters are within or close to the confidence intervals. 

We note that neither the estimated parameters nor the confidence intervals of the Monte Carlo method need to necessarily agree with those of the likelihood ratio method. 
First, while the Monte Carlo method is easier to implement, it is blind to systematic errors. For example, if the Baum-Welch algorithm repeatedly converges to an identical local (and not global) LL maximum, the confidence intervals will probably be severely underestimated. Generally speaking, the Monte Carlo method is a proxy to get confidence intervals that is simple, but not guaranteed in any sense. In contrast, the likelihood ratio method is, in principle, exact, were it implemented by multidimensional integration of the full LL. This, however, is most often not feasible and thus has to be replaced by the approximation given in Eq.~\eqref{eq:logLratio_eq}, the assumptions for which might be violated. We implement both methods to increase the chance that we spot any problematic behavior of the Baum-Welch algoritm. 

To analyze the reliability of the confidence intervals, we counted the number of times the true parameter is outside the $[\bw{\lambda}_0-3\delta_L^-,\bw{\lambda}_0+3\delta_L^+]$ or $[\bar{\lambda}_0-3\delta_\text{MC},\bar{\lambda}_0+3\delta_\text{MC}]$ interval.
If the likelihood was Gaussian shaped, those intervals would correspond to a 99.7\% confidence level. %
For the confidence interval calculations presented here, this occurs once, namely, for the MC confidence intervals of the Elzerman parameter $A_{13}$ at SNR $=5.0$, $A_{12}=A_{23}=0.001$.  We observed a similar behavior for other Elzerman readout calculations, again for the parameter $A_{13}$. The parameter corresponds to a direct transition from spin up to spin down without a change in the signal, which makes this parameter difficult to estimate. This difficulty seems to be reflected in a relatively flat LL landscape, correctly probed by the likelihood ratio method. However, the Monte Carlo method seems to repeatedly converge to the same value in parameter space and thereby underestimates the underlying uncertainty.

\section{Generation of correlated signals} \label{sec:generate_correlated_noise}
For completeness, here we review the method to generate a signal trace with correlated noise from Ref.~\cite{GutierrezRubio2022}. 
To generate a signal trace $y_{t}$ from a given noise spectrum $\Lambda_k$ and signal mean $\mu$, the values $\mathcal{R}y_k$ and $\mathcal{I}y_k$ are sampled for each $0\leq k\leq T-1$ from the distributions
\begin{align}
P(\mathcal{R}y_k|\Lambda_k,\mu) &= \sqrt{\frac{d_k}{\pi\lambda_k}}\exp\left(-d_k\frac{(\mathcal{R}y_k-\delta_{k,0}\sqrt{T}\mu)^2}{\Lambda_k}\right),\nonumber\\
P(\mathcal{I}y_k|\Lambda_k,\mu) &=
\begin{cases}
\delta_{\mathcal{I}y_k,0}\qquad\qquad\text{for }k=0 \text{ or }\lfloor{T/2}\rfloor,\\
\sqrt{\frac{d_k}{\pi\lambda_k}}\exp\left(-d_k\frac{(\mathcal{I}y_k)^2}{\Lambda_k}\right) \qquad\text{ else}.
\end{cases}
\end{align}
Here, $d_0=1/2$, $d_{T/2}=1/2$ (only for even $T$), and $d_k=1$ for all other $k$. The signal trace $y_{t}$ is obtained by taking the Fourier transform of $y_k$ = $\mathcal{R}y_k + i \mathcal{I}y_k$ via
\begin{align} \label{eq:signal_FT}
y_t = \frac{1}{\sqrt{T}}\sum_{k=0}^{T-1}y_k\exp\left(i2\pi tk/T\right).
\end{align}
The process is repeated for each signal trace $n$ of the measurement. We show some examples of the Gaussian-shaped spectrum and autocorrelation functions in Fig.~\ref{fig:gaussianfrequency_correlation}.

\begin{figure*}
(a)\includegraphics[width=0.4\textwidth]{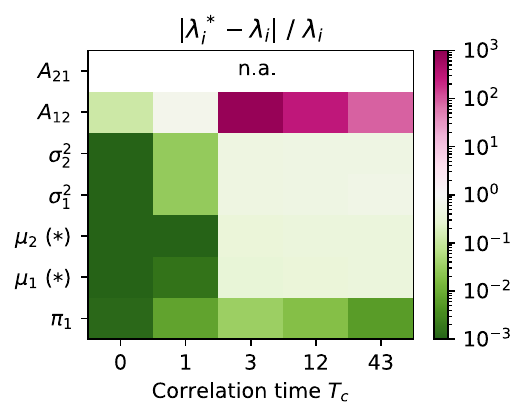}
(b)\includegraphics[width=0.4\textwidth]{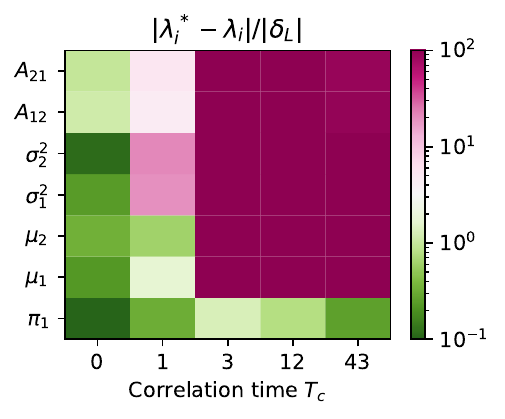}
\caption{The output of the Baum-Welch algorithm to estimate HMM parameters on data with correlated noise for the five noise profiles shown in Fig.~\ref{fig:gaussianfrequency_correlation}. We plot the difference between the estimated and model-matched parameters (a) relative to the amplitude of the model-matched parameters, and (b) relative to their confidence intervals as found by using Eq.~\eqref{eq:logLratio_eq}. We used $2000$ signal traces, each with $T=1000$ time steps, and chose $\Sigma_0=1$ and $A_{12}=0.0001$. We picked $\delta_L=\delta_L^+$ if $\lambda_i>\bw{\lambda}_i$, otherwise $\delta_L=\delta_L^-$. $(*)$ The parameters $\mu_1$ and $\mu_2$ are compared to their difference $|\mu_2-\mu_1|=1$ rather than their absolute value. 
}
\label{fig:parameter_estimation_correlation}
\end{figure*}

\section{Parameter estimation fails with correlated noise}
\label{sec:correlation_baumwelch}



We run the Baum-Welch algorithm on signals with correlated noise and calculate how the estimated parameters differ from their model-matched parameters. We show those deviations relative to the parameters' absolute values in Fig.~\ref{fig:parameter_estimation_correlation} (a) and to their likelihood ratio confidence intervals in Fig.~\ref{fig:parameter_estimation_correlation} (b). The correlated noise renders the parameter estimation unreliable; the estimated values and their confidence intervals strongly deviate from the model-matched values. The correlation time $T_c$ controls the amount of correlation in the signal trace, with $T_c=0$ corresponding to uncorrelated noise.  Already for $T_c=1$, the Baum-Welch estimated parameters become unreliable. Even when the relative error is small, the confidence intervals falsely suggest that the parameters have been estimated at a much higher precision. The confidence intervals of $\sigma_i^2$ are particularly underestimated, possibly because a clear relation to the autocorrelation function does not exist. %
The estimation of the relaxation probability $A_{12}$ also fails, which is unfortunate, because it is difficult to get the transition rates from other methods like the threshold method. 
The Baum-Welch algorithm was used before to extract the transition rates of electrons in quantum dots in Refs.~\cite{House2013, House2009}. Their results are reliable only if the noise correlations were negligibly small. We suggest that the Baum-Welch algorithm be tested on simulated data with the noise spectrum extracted from the experiment before it can be trusted as a reliable method to estimate physical parameters.


\bibliography{lit.bib}

\end{document}